\documentclass[aps,prl,twocolumn,superscriptaddress,showpacs]{revtex4-1}
\usepackage{graphics}
\usepackage{graphicx, bm}
\ifx\pdftexversion\undefined
\usepackage[dvips]{hyperref}
\else
\usepackage[colorlinks, allcolors=blue]{hyperref}
\fi
\hypersetup{
  colorlinks = true, linkcolor = blue
}

\begin{document}
\preprint{}

\title{Studying the low-entropy Mott transition of bosons in a three-dimensional optical lattice by measuring the full momentum-space density
}

\author{Ga\'etan Herc\'e}
\author{C\'ecile Carcy}
\author{Antoine Tenart}
\author{Jan-Philipp Bureik}
\author{Alexandre Dareau}
\author{David Cl\'ement}
\affiliation{Universit\'e Paris-Saclay, Institut d'Optique Graduate School, CNRS, Laboratoire Charles Fabry, 91127, Palaiseau, France}
\author{Tommaso Roscilde}
\affiliation{Universit\'e de Lyon, Ens de Lyon, Univ. Claude Bernard and CNRS, Laboratoire de Physique, F-69342 Lyon, France}
\date{\today}

\begin{abstract}
We report on a combined experimental and theoretical study of the low-entropy Mott transition for interacting bosons trapped in a three-dimensional (3D) cubic lattice -- namely, the interaction-induced superfluid-to-normal phase transition in the vicinity of the zero-temperature Mott transition. Our analysis relies on the measurement of the 3D momentum distribution, which allows us to extract the momentum-space density $\rho(\bm k=\bm 0)$ at the center of the Brillouin zone. Upon varying the ratio between the interaction $U$ and the tunnelling energy $J$ across the superfluid transition, we observe that $\rho(\bm k=\bm 0)$ exhibits a sharp transition at a value of $U/J$ consistent with the bulk prediction from quantum Monte Carlo. In addition, the variation of $\rho(\bm k=\bm 0)$ with $U/J$ exhibits a critical behavior consistent with the expected 3D XY universality class. Our results show that the tomographic reconstruction of the momentum distribution of ultracold bosons can reveal traits of the critical behavior of the superfluid transition even in an inhomogeneous trapped system. 
\end{abstract}

\keywords{}
\maketitle 

\emph{Introduction.} 
In condensed matter, the Mott transition is a celebrated metal-insulator transition induced by electron-electron Coulomb interactions \cite{Mott-book, transition-book}, and is found in a wide class of materials \cite{imada1998}. Over the past decades, the Mott transition has also become central in the field of quantum gases, as a paradigmatic example of an interaction-induced phase transition realized in experiments \cite{gross2017}. Cold-atom experiments can implement both the fermionic (metal-insulator) Mott transition \cite{jordens2008, schneider2008, hofrichter2016}, as well as its bosonic analog, in which a system of lattice bosons is driven from a superfluid (SF) Bose-Einstein condensate to an insulator \cite{greiner2002, mun2007, jimenez-garcia2010, becker2010, Trotzkyetal2010, mark2011, thomas2017} -- this latter transition will be the focus of our study. 

In the study of Mott physics, two important differences exist between experiments on solid-state materials and on quantum gases: 1)  while the electron density is essentially homogenous in a solid, the atomic density varies in a quantum gas as a result of the harmonic trap in which atoms are held; this difference complicates the analysis of cold-atom experiments, in particular when it comes to identifying the critical parameters of the Mott transition \cite{greiner2002, mun2007, jordens2008, schneider2008, jimenez-garcia2010, becker2010, mark2011, hofrichter2016, thomas2017}; and 2) while solid-state experiments are conducted in the presence of a heat bath fixing the temperature of the system, quantum-gas experiments are conducted at (nearly) constant entropy, and are typically subject to adiabatic heating \cite{Trotzkyetal2010,carcy2021}, so that they actually probe a finite-temperature (or finite-entropy) Mott transition. Both aspects suggest that it is rather challenging to observe the critical behavior of the transition in current quantum-gas experiments, particularly so in the vicinity of the zero-temperature quantum critical point. A motivation for the present work is to further elucidate these difficulties from a renewed perspective combining experiment and theory.

The low-entropy Mott transition in quantum-gas experiments can be explored by varying the $u=U/J$ ratio between the on-site interaction energy $U$ and the tunnelling energy $J$  for particles trapped in the lowest band of an optical lattice, realizing the physics of the single-band Hubbard model. In the case of three-dimensional (3D) lattice bosons -- to which we specialize our attention in the following -- the critical ratio $u_c$ for the appearance of the incompressible Mott insulator (MI) phase has been estimated in experiments with a variety of signatures:  by observing the appearance of a gap in the excitation spectrum \cite{greiner2002}; by observing  kinks in the visibility of the interference pattern \cite{gerbier2005} and in the width of the momentum distribution \cite{mark2011}; and by measuring the breakdown of SF currents \cite{mun2007}. Surprisingly, the values of the critical ratio $u_{c}$ estimated in several experiments were found to be compatible with the mean-field prediction \cite{mun2007, becker2010, thomas2017} for the homogeneous Bose-Hubbard model at zero temperature, rather than with its accurate estimate using quantum Monte Carlo (QMC)  (which is significantly lower than the MF value both at zero and at finite temperature) \cite{CapogrossoSansoneetal2007}; or unable to distinguish between the two predictions \cite{mark2011}. These results question the ability of quantum gas experiments to accurately locate the Mott transition.
 
In this work, we re-examine the experimental determination of the low-entropy Mott transition in 3D lattice bosons by using an original approach to this problem, namely the reconstruction of the full 3D momentum-space density $\rho(\bm k)$. Such an approach is implemented using the electronic detection method for metastable Helium atoms after time of flight \cite{cayla2018, tenart2020}. Previous experiments have relied on optical absorption imaging of atomic clouds after time of flight, which yields instead a 2D atomic density corresponding to the 3D density $\rho(\bm k)$ integrated along the line of sight. From the full measurement of $\rho(\bm k)$, on the other hand, one could in principle extract the condensate fraction $f_c$, namely the number of atoms in the in-trap condensate mode. Such a mode can be well identified in the case of bosons in shallow lattices, whose momentum-space density, similarly to the case of bosons without a lattice  \cite{ketterle1998}, exhibits a double structure (a condensate peak and a non-condensed pedestal). In Ref.~\cite{cayla2018} we used such an approach to investigate the BEC transition at a fixed value $u=10$. In contrast, in the vicinity of the Mott transition, the gas becomes strongly correlated, $f_{c}$ is small and a distinction between the condensed and non-condensed components is hardly possible from measuring the momentum-space density. The alternative we adopt in this work consists in monitoring the maximum  momentum-space density $\rho_0 = \rho({\bm k}={\bf 0})$ in the center of the Brillouin zone. This quantity is uniquely accessible via our reconstruction of the 3D atomic density. 

Since the condensate mode in momentum space is strongly peaked at $k=0$, the value of $\rho_0$ is intimately connected with the condensed fraction \cite{ma2008}. Most importantly, it is directly measured in the experiment without relying on any fit of the momentum-space density. Monitoring the dependence of $\rho_0$ on the ratio $u=U/J$, we observe a low-entropy Mott transition with one atom per site at the critical ratio $u_{c}=26(1)$. Building {\color{blue} on} accurate thermometry based on a systematic comparison with ab-initio QMC data \cite{carcy2021}, we show that this estimate is consistent with what is expected for the homogeneous Bose-Hubbard model at the same temperature as that of the experiment in the critical regime. As we shall discuss below, such an agreement is not generic, and it strongly depends on the specific conditions of our experiment. More interestingly, we conduct an analysis of the critical suppression of the value $\rho_0$ on the SF side of the transition, as well as of the critical shrinking of the momentum peak width on the normal side. We find that the first quantity reveals the expected critical behavior of the superfluid-to-normal transition of a bulk system, while the second is more strongly affected by the trap.   
  
 \begin{figure}[ht!]
\includegraphics[width=\columnwidth]{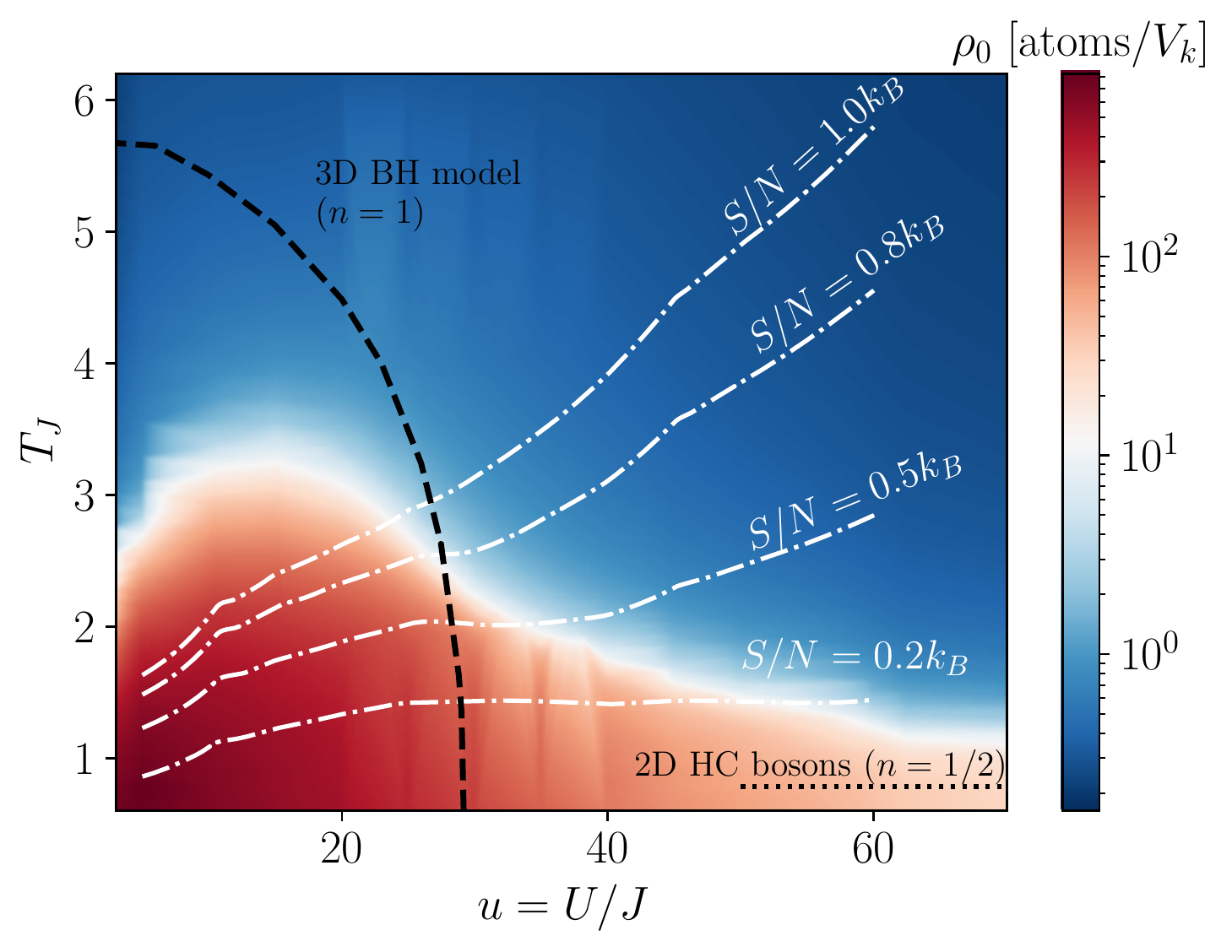}
\caption{{\bf Phase diagram of interacting 3D lattice bosons in a trap.} Numerical calculation of the momentum-space density $\rho_0$ at the center of the 3D Brillouin zone in the plane $u-T_{J}$, with $u=U/J$ and $T_{J}=k_{B}T/J$. $\rho_0$ is expressed in units of the atom number in a cube of volume $V_{k}=(k_{d}/30)^3$. The quantity $\rho_0$, plotted in false color, is obtained from Quantum Monte-Carlo calculations for 3D lattice bosons in a trap with the parameters of the experiment (see main text). The solid lines are isentropic curves obtained from the QMC calculations, for the same experimental parameters \cite{carcy2021}. The black dashed line is the location of the superfluid-to-normal transition in the homogeneous 3D Bose-Hubbard model with unit filling $n=1$ \cite{CapogrossoSansoneetal2007}. The dotted line indicates the critical temperature $k_{B}T_{c}\simeq0.785J/k_{B}$ for 2D hard-core (HC) bosons at half filling $n=1/2$ \cite{HaradaK1998}.
}
\label{fig1}
\end{figure} 
  
Our experiment aims at the quantum simulation of the 3D Bose-Hubbard (BH) model in a harmonic trap,
\begin{equation}
{\cal H} = - J\sum_{\langle ij \rangle} \left ( b_i^\dagger b_j + {\rm h.c.} \right ) + \frac{U}{2} \sum_i n_i (n_i-1) + \sum_i V_i n_i,
\label{e.BH}
\end{equation} 
where $\langle ij \rangle$ denotes a pair of nearest-neighbor sites, $J$ and $U$ have been introduced above, and $V_i = (1/2) m\omega^2 r^2_i$ is an overall harmonic trapping potential at the position ${\bm r}_i$ of the $i$-th site, $m$ is the atom mass and $\omega/2\pi$ the trapping frequency.  
We investigate the low-entropy Mott transition of the Bose-Hubbard model with a gas of metastable Helium-4 ($^4$He$^*$) atoms as described in \cite{carcy2021}. In brief, we load a Bose-Einstein condensate of $N=3000(400)$ $^4$He$^*$ atoms in a 3D optical lattice with a lattice spacing of $d=775~$nm. The choice of the particle number gives a density at the center of the trap of $n\lesssim 1$ atoms per site. We measure the 3D distribution of individual atoms in momentum space \cite{cayla2018} at various amplitudes of the lattice, corresponding to different $u=U/J$ ratios. From these distributions, we extract the momentum-space density $\rho(\bm k)$. In addition, a direct comparison of the measured $\rho(\bm k)/\rho_{0}$ with that predicted by ab-initio QMC calculations with the temperature as the only adjustable parameter provides us with an accurate thermometry for the lattice gas \cite{carcy2021}.

\emph{ QMC data for the trapped system.} 
When considering the BH model at filling $n=1$ in the absence of a trap, one expects a SF phase at low $u$ and low temperature up to a critical temperature $T_c$. At zero temperature, the SF phase terminates at the Mott quantum-critical point at $u_c^{(n=1)} = 29.3$ \cite{CapogrossoSansoneetal2007}. At finite temperature, it terminates at the critical temperature $T_c(u)$, which is reported in Fig.~\ref{fig1} as a dashed line in the case of the homogeneous BH model. In Fig.~\ref{fig1}, we also plot the theoretical values $\rho_{0}$ obtained from extensive {\color{blue} QMC}  data for the BH model in a trap with $N=3000$ atoms and trapping potential as in the experiment, over the intervals of $u$ and  $T_J=k_{B} T/J$ ratios that are relevant for our experiment. From the value of $\rho_0$ one can clearly identify a SF regime ($\rho_0 \gtrsim 10$ atoms/$V_k$  where $V_k = (k_d/30)^3$), whose temperature range is slightly increased with $u$ when $u\lesssim 15$ (because the trapping potential is establishing a density $n<1$ in the trap center in this range), and which is instead systematically suppressed with increasing $u$ above $u = 15$. This latter trend is to be expected from the fact that atoms in the trap center develop a MI phase for $u > u_c^{(n=1)}$, and therefore massively disappear from the condensate peak. At the same time, atoms in the outer halo of the atomic cloud remain superfluid at sufficiently low temperatures for all interaction values. As $u \to \infty$ the halo shrinks to a thin spherical corona with average density $n=1/2$ atoms per site, reproducing the physics of 2D hardcore bosons at half filling, which exhibit a 2D SF transition at a temperature $T_c \approx 0.785 J/k_{B}$ \cite{HaradaK1998}. This implies that, when looking at global coherence properties in this regime of ultra-low temperatures, the physics of the $T=0$ Mott transition is significantly masked by the persistence of superfluidity in the outer halo. 

Fig.~\ref{fig1} also shows theoretical isentropic curves in the $u-T_J$ plane. In Ref.~\cite{carcy2021} we established that in our experiment atoms are adiabatically loaded into the optical lattice, so that the experiment follows approximately an isentropic line with $S/N = 0.8 k_B$ -- an entropy corresponding to that of the Bose-Einstein condensate before it is loaded into the lattice. The evolution of $\rho_{0}$ along this isentropic line  highlights two features of the transition which we probe in the experiment. On the one hand, a (finite-size) superfluid-to-normal transition is expected at a finite temperature $T_{\rm exp} \approx 2.6 J$ and at an interaction $u_c(T_{\rm exp})$ very close to the bulk value $u_c^{(n=1)}$ for the $T=0$ Mott transition. On the other hand, the fact that $u_c(T_{\rm exp}) < u_c^{(n=1)}$ implies that the transition in the trapped system is driven by the loss of coherence in the trap center developing a finite-$T$ Mott phase, and does not involve the halo (which is fully normal at $T_{\rm exp}$).  All these aspects put together allow us to qualify the transition that we explore experimentally as a {\it low-entropy Mott transition}. Nonetheless, we expect it to exhibit critical properties of the finite-temperature superfluid-to-normal transition, and not that of the $T=0$ Mott transition.
 
\begin{figure}[ht!]
\includegraphics[width=\columnwidth]{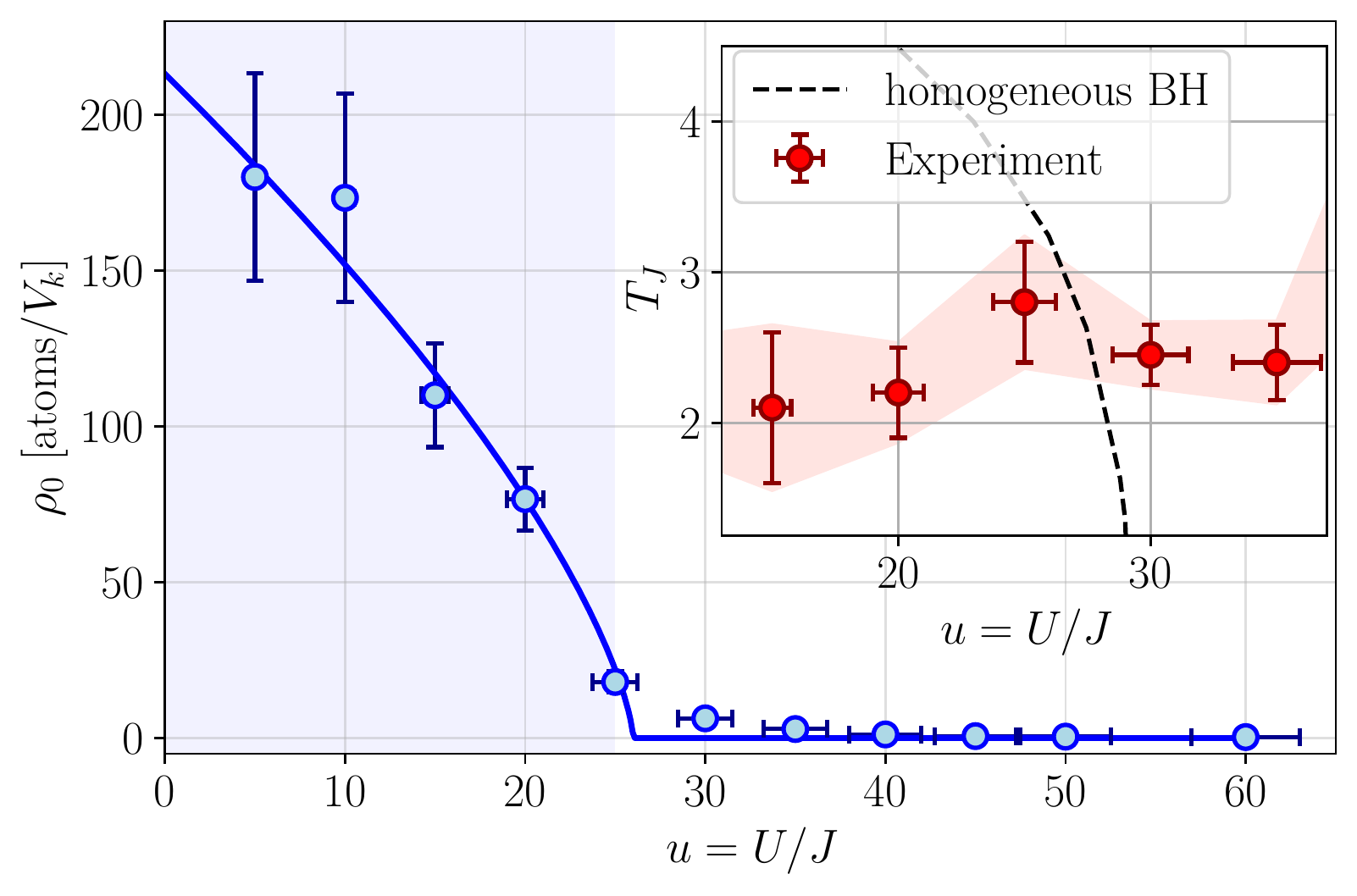}
\caption{{\bf Identifying the Mott transition.} Plot of the central momentum-space density $\rho_0$ as a function of $u=U/J$. The solid blue line is a fit of the experimental data for $u \leq 25$ (shaded blue region) with the function $\rho_{0}^{u=0}(1-u/u_{c})^{2 \beta}$ where $\beta=0.3485$, while $\rho_{0}^{u=0}$ and $u_{c}$ are fitting parameters. Inset: from the experimental temperatures \cite{carcy2021} we extract the corresponding critical ratio $u$ for the homogeneous BH model via the knowledge of the $u_c=u_c(T)$ curve (dashed line \cite{CapogrossoSansoneetal2007}).}
\label{fig2}
\end{figure}
 
\emph{Critical ratio $u_{c}$ and critical behavior of $\rho_0$.} 
A detailed insight into the critical behavior of the momentum-space density at the Mott transition is offered by the dependence of $\rho_0$ with $u$, as shown in Fig.~\ref{fig2}. We observe that $\rho_0$ drops rapidly with $u$ in the range $u \leq 25$, thereby exhibiting a sharp transition in spite of the inhomogeneous nature of our system. Finite-size effects can be clearly seen in the appearance of a tail for $u\geq 30$. The error bars, due to the shot noise in the detection process, are relatively large here because we use a rather low detection efficiency (of $5\%$) in this work to avoid saturating the He$^*$ detector. In order to extract $u_c$ from these data we fit them in the window $u \leq 25$ with the behavior expected to hold close to the critical point in the homogeneous case, namely $\rho_0(u) = \rho_{0}^{u=0} |1-u/u_c|^{2 \beta}$. Here $\rho_{0}^{u=0}$ and $u_c$ are fitting parameters, while we take $\beta=0.3485$ \cite{campostrini2001} as expected for the 3D XY universality class. The fit is rather convincing, with resulting fitting parameters $\rho_{0}^{u=0}=215(15)~$atoms$/V_{k}$ and $u_{c}=26(1)$. The value of the critical interaction strength is consistent with that of the homogeneous BH model at density $n=1$ at the temperature estimated for the experiment,  $u_c = 27(1)$ (see the inset in Fig.~\ref{fig2}). This value is clearly incompatible with the zero-temperature mean-field prediction, $u_c = 34.5$. Our observations further indicate that the transition we are observing stems from the formation of a Mott-insulating core in the center of the trap, and not from a loss of coherence in the cloud halo. 

\begin{figure}[h!]
\includegraphics[width=\columnwidth]{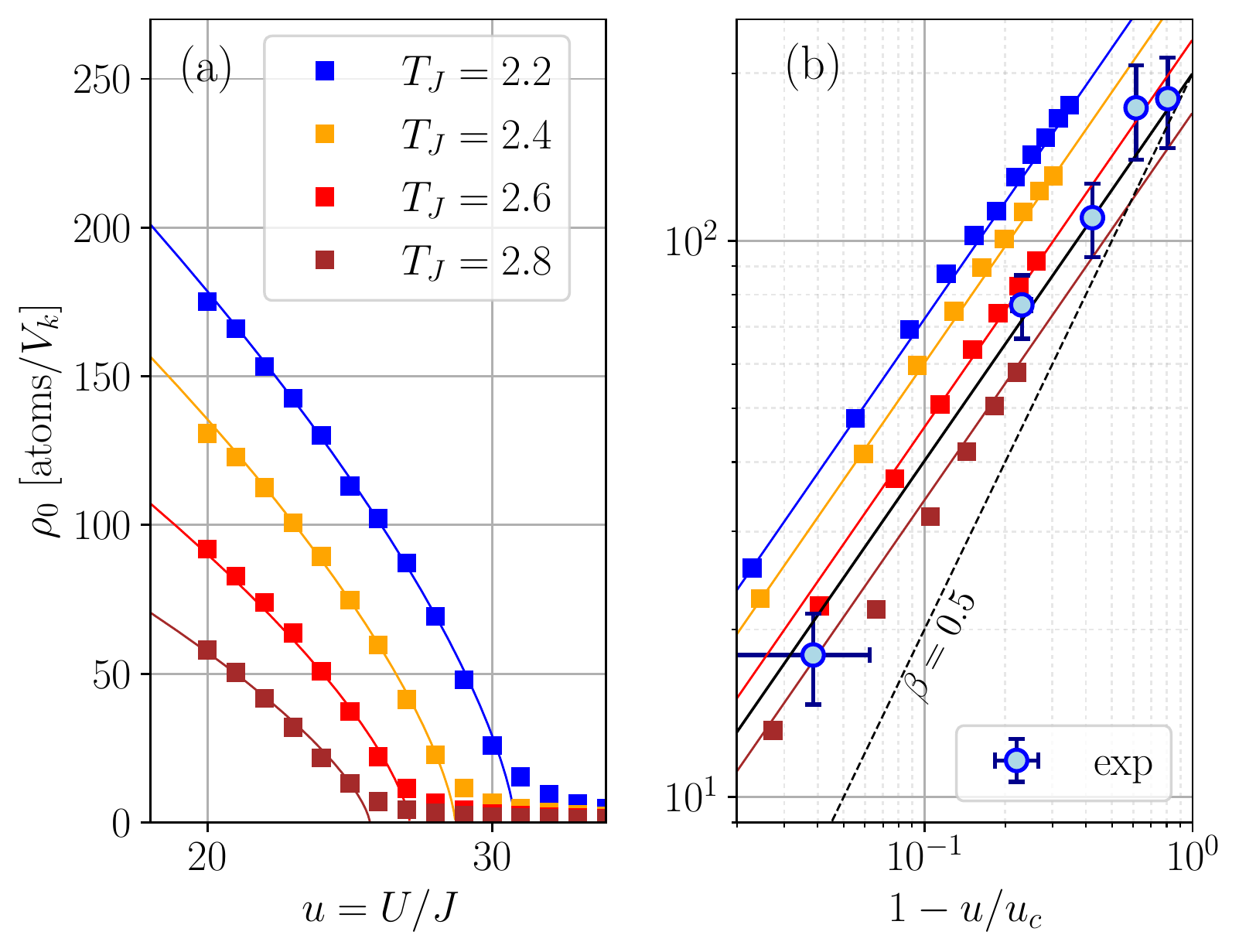}
\caption{{\bf Critical behavior from QMC data and experiments.} (a) QMC data for $\rho_0$ along isotherms at temperatures close to the experimental one. The solid lines are fits to $\rho_0^{u=0}(1-u/u_c)^{2\beta}$, with $u_c$ representing an effective critical coupling for the trapped system (deviating from the $u_c$ value of the homogeneous system at unit filling); (b) Log-log plot of the same data, along with the experimental ones, showing the range of critical behavior. All solid lines are fit to the QMC or experimental data with the exponent $\beta=0.3485$ of the 3D XY universality class. The dashed line is the expected mean-field critical behavior for which $\beta=0.5$ (see main text).}
\label{fig3}
\end{figure}

It may appear surprising at first sight that the critical behavior of the bulk system is observable in a trapped system, and that it is apparently manifested in a rather broad range of interactions $u$ below the critical point. To corroborate this observation, we compare in Fig.~\ref{fig3} the experimental data with QMC data for trapped bosons obtained along isotherms at temperatures close to $T_{\rm exp}$. Note that, even though the experimental data follow strictly speaking an isentropic curve, the temperature along this curve shows only a moderate variation around $T_{\rm exp}$, so that the comparison with isotherms is meaningful. There, we observe that the bulk critical behavior is indeed exhibited by the QMC data for the trapped system over a rather broad range of interaction values; and that both the experimental and the theoretical data are clearly incompatible with the mean-field criticality (namely with $\beta = 1/2$). We note, however, that the effective critical value $u_{c}$ of the trapped system leading to the observation of the critical behavior generally differs from that of the bulk. This is illustrated by the critical value of the trapped QMC data at $T_{J}=2.2$ ($u_{c}\sim31$) which exceeds the largest critical ratio $u_{c}$ of the homogeneous case (see Fig.~\ref{fig3}a) because of the contribution coming from the cloud halo.

\begin{figure}[h!]
\includegraphics[width=\columnwidth]{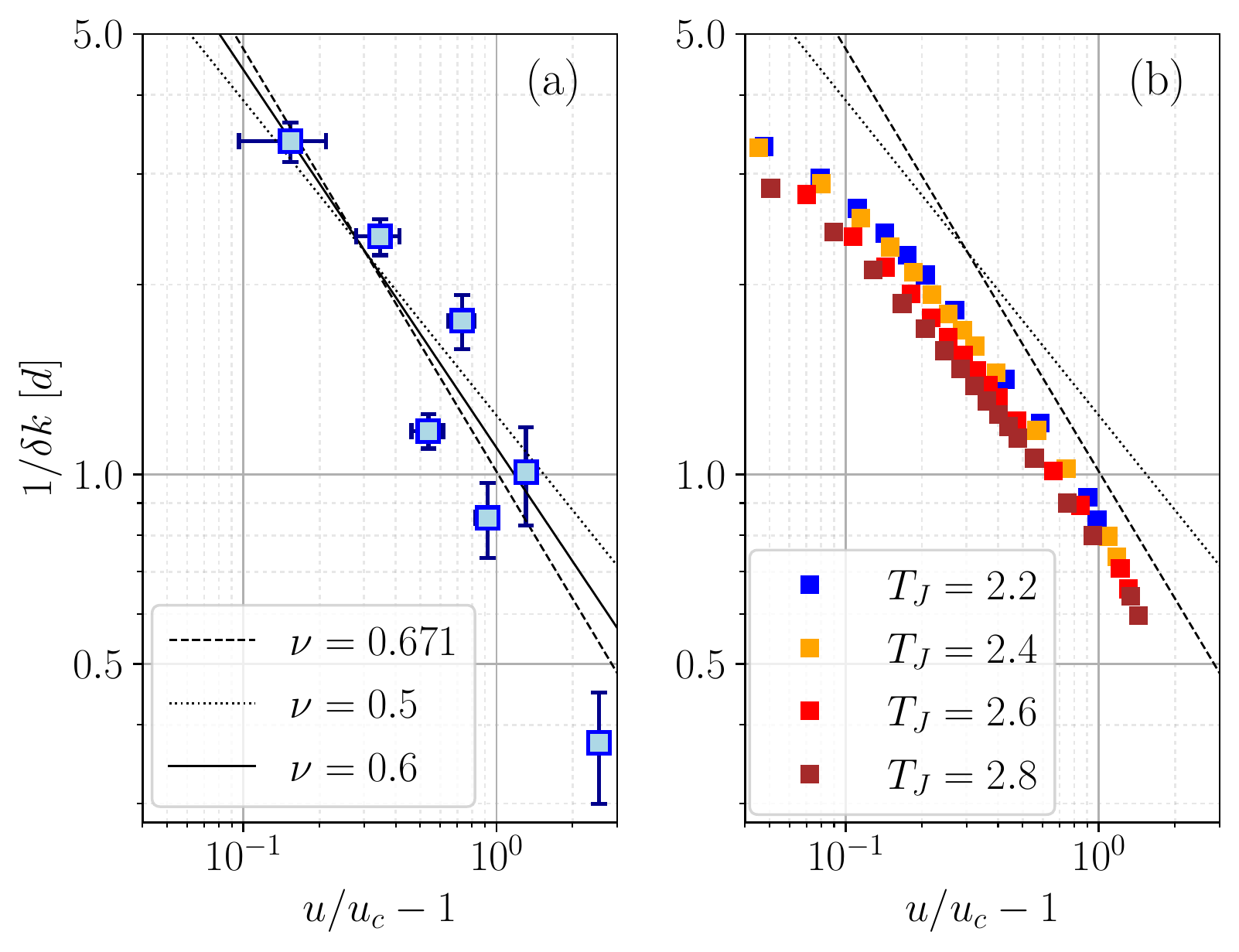}
\caption{{\bf Phase coherence properties.} (a) Log-log plot of the inverse of HWHM $\delta k$ of $\rho({\bf k})$ in momentum space as a function of $u/u_c-1$. The solid line is a fit $\xi_0 \times (u/u_c-1)^\nu$ to the data (with fitting parameters $\xi_0$ and $\nu$), while the dashed and the dotted lines are fits of $\xi_0$ with fixed $\nu = 0.671$ and $\nu = 0.5$ respectively. Only the experimental data in the range $u \in [30;60]$ is used for fitting. (b) QMC data for the same quantity along isotherms close to $T_{\rm exp}$; while an apparent critical behavior is observed in the data, it is rather consistent with mean-field criticality ($\nu =1/2$, dotted line) than with the 3DXY one ($\nu = 0.671$, dashed line).}
\label{fig4}
\end{figure}

\emph{Correlation length from the HWHM.} 
Further insight into the critical behavior of the momentum-space density can be gained by examining the insulator region for values $u>u_{c}$. To this aim, we extract the half width at half maximum (HWHM) $\delta k$ of the momentum-space density.   The width $\delta k$ provides information about the spatial coherence of the lattice gas, as it is inversely proportional to the in-trap phase-coherence length $l_{\phi}$. In a homogeneous system, upon approaching the phase transition from the MI regime  $l_{\phi}$ is expected to increase as $l_{\phi} \propto (u-u_{c})^{-\nu}$, with $\nu=0.671$ (from the 3DXY universality class \cite{campostrini2001}). At the mean-field level, a similar critical behavior is expected for $\delta k(u)$, but with an exponent $\nu=1/2$. Fig.~\ref{fig4} shows that a critical behavior of the kind $l_{\phi} \propto (u-u_{c})^{-\nu}$ is found to be compatible with the experimental data. But the error bars and the scatter of the experimental point are rather significant, and we are not able to discriminate between the mean-field prediction and that of the 3DXY universality class. Indeed, when fitting $1/\delta k$ to $\xi_0 (u/u_c-1)^{-\nu}$ with $\xi_0$ and $\nu$ as fitting parameters, we obtain a value $\nu=0.6(1)$. Moreover the experimental data follow an isentropic curve which is a rather peculiar trajectory approaching the critical point, and in particular one that deviates significantly from an isotherm in the interaction range $u \gtrsim 40$ (see Fig.~\ref{fig1}). Nonetheless, we also show in Fig.~\ref{fig4}(b) our QMC data for $1/\delta k$ (upon approaching the critical point along isotherms). These results confirm that $1/\delta k$ indeed exhibits an apparent critical behavior, but one which is rather compatible with the mean-field exponent. On the basis of these numerical data we conclude that the trap effects in our experiment are too strong to clearly observe the bulk critical behavior of the correlation length on the MI side. 

\emph{Conclusions.} 
In this work we have provided a detailed study of the low-entropy Mott transition of 3D trapped lattice bosons, probed via the full 3D momentum-space density $\rho(\bm k)$. Focusing on the interaction dependence of $\rho_0 = \rho(\bm k=0)$, we estimated the critical interaction strength, which turns out to be in agreement with the expected QMC value from the finite-temperature transition of the uniform Bose-Hubbard model at unit filling. In spite of the presence of the trap, for $u$ below the critical coupling the experimental data show a critical suppression of $\rho_0$ which is consistent with the expected behavior in the bulk system, exposing a non-mean-field critical exponent. On the other hand the observation of the expected critical behavior for the correlation length on the normal side of the transition is strongly hindered by the trap. Our results show that the presence of the harmonic trap, while serving the role of fixing the density at $n=1$ upon approaching the Mott insulator phase, complicates the observation of the critical behavior of the Mott transition. Most importantly, our QMC data show that the presence of a superfluid halo around the cloud core would prevent the observation of quantum criticality when reducing the entropy significantly below the value attained by our experiment. One could circumvent this difficulty by using homogenous traps \cite{Gaunt2013} or by selectively probing the coherence properties at the center of the trap \cite{Sagietal2012}.

\vspace{0.5cm}
\begin{acknowledgments}
We acknowledge fruitful discussions with A. Ran\c con and all the members of the Quantum Gas group at Institut d'Optique. We acknowledge financial support from the LabEx PALM (Grant number ANR-10-LABX-0039), the R\'egion Ile-de-France in the framework of the DIM SIRTEQ, the ``Fondation d'entreprise iXcore pour la Recherche", the Agence Nationale pour la Recherche (Grant number ANR-17-CE30-0020-01). D.C. acknowledges support from the Institut Universitaire de France. All the numerical simulations were performed on the PSMN cluster at the ENS of Lyon. 
\end{acknowledgments}

\bibliography{SFMI_3DBH}

\begin{thebibliography}{25}%
\makeatletter
\providecommand \@ifxundefined [1]{%
 \@ifx{#1\undefined}
}%
\providecommand \@ifnum [1]{%
 \ifnum #1\expandafter \@firstoftwo
 \else \expandafter \@secondoftwo
 \fi
}%
\providecommand \@ifx [1]{%
 \ifx #1\expandafter \@firstoftwo
 \else \expandafter \@secondoftwo
 \fi
}%
\providecommand \natexlab [1]{#1}%
\providecommand \enquote  [1]{``#1''}%
\providecommand \bibnamefont  [1]{#1}%
\providecommand \bibfnamefont [1]{#1}%
\providecommand \citenamefont [1]{#1}%
\providecommand \href@noop [0]{\@secondoftwo}%
\providecommand \href [0]{\begingroup \@sanitize@url \@href}%
\providecommand \@href[1]{\@@startlink{#1}\@@href}%
\providecommand \@@href[1]{\endgroup#1\@@endlink}%
\providecommand \@sanitize@url [0]{\catcode `\\12\catcode `\$12\catcode
  `\&12\catcode `\#12\catcode `\^12\catcode `\_12\catcode `\%12\relax}%
\providecommand \@@startlink[1]{}%
\providecommand \@@endlink[0]{}%
\providecommand \url  [0]{\begingroup\@sanitize@url \@url }%
\providecommand \@url [1]{\endgroup\@href {#1}{\urlprefix }}%
\providecommand \urlprefix  [0]{URL }%
\providecommand \Eprint [0]{\href }%
\providecommand \doibase [0]{http://dx.doi.org/}%
\providecommand \selectlanguage [0]{\@gobble}%
\providecommand \bibinfo  [0]{\@secondoftwo}%
\providecommand \bibfield  [0]{\@secondoftwo}%
\providecommand \translation [1]{[#1]}%
\providecommand \BibitemOpen [0]{}%
\providecommand \bibitemStop [0]{}%
\providecommand \bibitemNoStop [0]{.\EOS\space}%
\providecommand \EOS [0]{\spacefactor3000\relax}%
\providecommand \BibitemShut  [1]{\csname bibitem#1\endcsname}%
\let\auto@bib@innerbib\@empty
\bibitem [{\citenamefont {Mott}(2004)}]{Mott-book}%
  \BibitemOpen
  \bibfield  {author} {\bibinfo {author} {\bibfnamefont {N.}~\bibnamefont
  {Mott}},\ }\href@noop {} {\emph {\bibinfo {title} {Metal-Insulator
  Transitions}}}\ (\bibinfo  {publisher} {CRC Press},\ \bibinfo {year}
  {2004})\BibitemShut {NoStop}%
\bibitem [{\citenamefont {Dobrosavljevic}\ \emph {et~al.}(2012)\citenamefont
  {Dobrosavljevic}, \citenamefont {Trivedi},\ and\ \citenamefont
  {Valles~Jr}}]{transition-book}%
  \BibitemOpen
  \bibinfo {editor} {\bibfnamefont {V.}~\bibnamefont {Dobrosavljevic}},
  \bibinfo {editor} {\bibfnamefont {N.}~\bibnamefont {Trivedi}}, \ and\
  \bibinfo {editor} {\bibfnamefont {J.}~\bibnamefont {Valles~Jr}},\ eds.,\
  \href@noop {} {\emph {\bibinfo {title} {Conductor-Insulator Quantum Phase
  Transitions}}}\ (\bibinfo  {publisher} {Cambridge},\ \bibinfo {year}
  {2012})\BibitemShut {NoStop}%
\bibitem [{\citenamefont {Imada}\ \emph {et~al.}(1998)\citenamefont {Imada},
  \citenamefont {Fujimori},\ and\ \citenamefont {Tokura}}]{imada1998}%
  \BibitemOpen
  \bibfield  {author} {\bibinfo {author} {\bibfnamefont {M.}~\bibnamefont
  {Imada}}, \bibinfo {author} {\bibfnamefont {A.}~\bibnamefont {Fujimori}}, \
  and\ \bibinfo {author} {\bibfnamefont {Y.}~\bibnamefont {Tokura}},\ }\href
  {\doibase 10.1103/RevModPhys.70.1039} {\bibfield  {journal} {\bibinfo
  {journal} {Rev. Mod. Phys.}\ }\textbf {\bibinfo {volume} {70}},\ \bibinfo
  {pages} {1039} (\bibinfo {year} {1998})}\BibitemShut {NoStop}%
\bibitem [{\citenamefont {Gross}\ and\ \citenamefont
  {Bloch}(2017)}]{gross2017}%
  \BibitemOpen
  \bibfield  {author} {\bibinfo {author} {\bibfnamefont {C.}~\bibnamefont
  {Gross}}\ and\ \bibinfo {author} {\bibfnamefont {I.}~\bibnamefont {Bloch}},\
  }\href {\doibase 10.1126/science.aal3837} {\ \textbf {\bibinfo {volume}
  {357}},\ \bibinfo {pages} {995} (\bibinfo {year} {2017})}\BibitemShut
  {NoStop}%
\bibitem [{\citenamefont {J{\"o}rdens}\ \emph {et~al.}(2008)\citenamefont
  {J{\"o}rdens}, \citenamefont {Strohmaier}, \citenamefont {G{\"u}nter},
  \citenamefont {Moritz},\ and\ \citenamefont {Esslinger}}]{jordens2008}%
  \BibitemOpen
  \bibfield  {author} {\bibinfo {author} {\bibfnamefont {R.}~\bibnamefont
  {J{\"o}rdens}}, \bibinfo {author} {\bibfnamefont {N.}~\bibnamefont
  {Strohmaier}}, \bibinfo {author} {\bibfnamefont {K.}~\bibnamefont
  {G{\"u}nter}}, \bibinfo {author} {\bibfnamefont {H.}~\bibnamefont {Moritz}},
  \ and\ \bibinfo {author} {\bibfnamefont {T.}~\bibnamefont {Esslinger}},\
  }\href {\doibase 10.1038/nature07244} {\bibfield  {journal} {\bibinfo
  {journal} {Nature}\ }\textbf {\bibinfo {volume} {455}},\ \bibinfo {pages}
  {204} (\bibinfo {year} {2008})}\BibitemShut {NoStop}%
\bibitem [{\citenamefont {Schneider}\ \emph {et~al.}(2008)\citenamefont
  {Schneider}, \citenamefont {Hackerm{\"u}ller}, \citenamefont {Will},
  \citenamefont {Best}, \citenamefont {Bloch}, \citenamefont {Costi},
  \citenamefont {Helmes}, \citenamefont {Rasch},\ and\ \citenamefont
  {Rosch}}]{schneider2008}%
  \BibitemOpen
  \bibfield  {author} {\bibinfo {author} {\bibfnamefont {U.}~\bibnamefont
  {Schneider}}, \bibinfo {author} {\bibfnamefont {L.}~\bibnamefont
  {Hackerm{\"u}ller}}, \bibinfo {author} {\bibfnamefont {S.}~\bibnamefont
  {Will}}, \bibinfo {author} {\bibfnamefont {T.}~\bibnamefont {Best}}, \bibinfo
  {author} {\bibfnamefont {I.}~\bibnamefont {Bloch}}, \bibinfo {author}
  {\bibfnamefont {T.~A.}\ \bibnamefont {Costi}}, \bibinfo {author}
  {\bibfnamefont {R.~W.}\ \bibnamefont {Helmes}}, \bibinfo {author}
  {\bibfnamefont {D.}~\bibnamefont {Rasch}}, \ and\ \bibinfo {author}
  {\bibfnamefont {A.}~\bibnamefont {Rosch}},\ }\href {\doibase
  10.1126/science.1165449} {\ \textbf {\bibinfo {volume} {322}},\ \bibinfo
  {pages} {1520} (\bibinfo {year} {2008})}\BibitemShut {NoStop}%
\bibitem [{\citenamefont {Hofrichter}\ \emph {et~al.}(2016)\citenamefont
  {Hofrichter}, \citenamefont {Riegger}, \citenamefont {Scazza}, \citenamefont
  {H\"ofer}, \citenamefont {Fernandes}, \citenamefont {Bloch},\ and\
  \citenamefont {F\"olling}}]{hofrichter2016}%
  \BibitemOpen
  \bibfield  {author} {\bibinfo {author} {\bibfnamefont {C.}~\bibnamefont
  {Hofrichter}}, \bibinfo {author} {\bibfnamefont {L.}~\bibnamefont {Riegger}},
  \bibinfo {author} {\bibfnamefont {F.}~\bibnamefont {Scazza}}, \bibinfo
  {author} {\bibfnamefont {M.}~\bibnamefont {H\"ofer}}, \bibinfo {author}
  {\bibfnamefont {D.~R.}\ \bibnamefont {Fernandes}}, \bibinfo {author}
  {\bibfnamefont {I.}~\bibnamefont {Bloch}}, \ and\ \bibinfo {author}
  {\bibfnamefont {S.}~\bibnamefont {F\"olling}},\ }\href {\doibase
  10.1103/PhysRevX.6.021030} {\bibfield  {journal} {\bibinfo  {journal} {Phys.
  Rev. X}\ }\textbf {\bibinfo {volume} {6}},\ \bibinfo {pages} {021030}
  (\bibinfo {year} {2016})}\BibitemShut {NoStop}%
\bibitem [{\citenamefont {Greiner}\ \emph {et~al.}(2002)\citenamefont
  {Greiner}, \citenamefont {Mandel}, \citenamefont {Esslinger}, \citenamefont
  {H{\"a}nsch},\ and\ \citenamefont {Bloch}}]{greiner2002}%
  \BibitemOpen
  \bibfield  {author} {\bibinfo {author} {\bibfnamefont {M.}~\bibnamefont
  {Greiner}}, \bibinfo {author} {\bibfnamefont {O.}~\bibnamefont {Mandel}},
  \bibinfo {author} {\bibfnamefont {T.}~\bibnamefont {Esslinger}}, \bibinfo
  {author} {\bibfnamefont {T.~W.}\ \bibnamefont {H{\"a}nsch}}, \ and\ \bibinfo
  {author} {\bibfnamefont {I.}~\bibnamefont {Bloch}},\ }\href {\doibase
  10.1038/415039a} {\bibfield  {journal} {\bibinfo  {journal} {Nature}\
  }\textbf {\bibinfo {volume} {415}},\ \bibinfo {pages} {39} (\bibinfo {year}
  {2002})}\BibitemShut {NoStop}%
\bibitem [{\citenamefont {Mun}\ \emph {et~al.}(2007)\citenamefont {Mun},
  \citenamefont {Medley}, \citenamefont {Campbell}, \citenamefont {Marcassa},
  \citenamefont {Pritchard},\ and\ \citenamefont {Ketterle}}]{mun2007}%
  \BibitemOpen
  \bibfield  {author} {\bibinfo {author} {\bibfnamefont {J.}~\bibnamefont
  {Mun}}, \bibinfo {author} {\bibfnamefont {P.}~\bibnamefont {Medley}},
  \bibinfo {author} {\bibfnamefont {G.~K.}\ \bibnamefont {Campbell}}, \bibinfo
  {author} {\bibfnamefont {L.~G.}\ \bibnamefont {Marcassa}}, \bibinfo {author}
  {\bibfnamefont {D.~E.}\ \bibnamefont {Pritchard}}, \ and\ \bibinfo {author}
  {\bibfnamefont {W.}~\bibnamefont {Ketterle}},\ }\href {\doibase
  10.1103/PhysRevLett.99.150604} {\bibfield  {journal} {\bibinfo  {journal}
  {Phys. Rev. Lett.}\ }\textbf {\bibinfo {volume} {99}},\ \bibinfo {pages}
  {150604} (\bibinfo {year} {2007})}\BibitemShut {NoStop}%
\bibitem [{\citenamefont {Jim\'enez-Garc\'{\i}a}\ \emph
  {et~al.}(2010)\citenamefont {Jim\'enez-Garc\'{\i}a}, \citenamefont {Compton},
  \citenamefont {Lin}, \citenamefont {Phillips}, \citenamefont {Porto},\ and\
  \citenamefont {Spielman}}]{jimenez-garcia2010}%
  \BibitemOpen
  \bibfield  {author} {\bibinfo {author} {\bibfnamefont {K.}~\bibnamefont
  {Jim\'enez-Garc\'{\i}a}}, \bibinfo {author} {\bibfnamefont {R.~L.}\
  \bibnamefont {Compton}}, \bibinfo {author} {\bibfnamefont {Y.-J.}\
  \bibnamefont {Lin}}, \bibinfo {author} {\bibfnamefont {W.~D.}\ \bibnamefont
  {Phillips}}, \bibinfo {author} {\bibfnamefont {J.~V.}\ \bibnamefont {Porto}},
  \ and\ \bibinfo {author} {\bibfnamefont {I.~B.}\ \bibnamefont {Spielman}},\
  }\href {\doibase 10.1103/PhysRevLett.105.110401} {\bibfield  {journal}
  {\bibinfo  {journal} {Phys. Rev. Lett.}\ }\textbf {\bibinfo {volume} {105}},\
  \bibinfo {pages} {110401} (\bibinfo {year} {2010})}\BibitemShut {NoStop}%
\bibitem [{\citenamefont {Becker}\ \emph {et~al.}(2010)\citenamefont {Becker},
  \citenamefont {Soltan-Panahi}, \citenamefont {Kronjäger}, \citenamefont
  {Dörscher}, \citenamefont {Bongs},\ and\ \citenamefont
  {Sengstock}}]{becker2010}%
  \BibitemOpen
  \bibfield  {author} {\bibinfo {author} {\bibfnamefont {C.}~\bibnamefont
  {Becker}}, \bibinfo {author} {\bibfnamefont {P.}~\bibnamefont
  {Soltan-Panahi}}, \bibinfo {author} {\bibfnamefont {J.}~\bibnamefont
  {Kronjäger}}, \bibinfo {author} {\bibfnamefont {S.}~\bibnamefont
  {Dörscher}}, \bibinfo {author} {\bibfnamefont {K.}~\bibnamefont {Bongs}}, \
  and\ \bibinfo {author} {\bibfnamefont {K.}~\bibnamefont {Sengstock}},\ }\href
  {\doibase 10.1088/1367-2630/12/6/065025} {\bibfield  {journal} {\bibinfo
  {journal} {New Journal of Physics}\ }\textbf {\bibinfo {volume} {12}},\
  \bibinfo {pages} {065025} (\bibinfo {year} {2010})}\BibitemShut {NoStop}%
\bibitem [{\citenamefont {Trotzky}\ \emph {et~al.}(2010)\citenamefont
  {Trotzky}, \citenamefont {Pollet}, \citenamefont {Gerbier}, \citenamefont
  {Schnorrberger}, \citenamefont {Bloch}, \citenamefont {Prokof'ev},
  \citenamefont {Svistunov},\ and\ \citenamefont {Troyer}}]{Trotzkyetal2010}%
  \BibitemOpen
  \bibfield  {author} {\bibinfo {author} {\bibfnamefont {S.}~\bibnamefont
  {Trotzky}}, \bibinfo {author} {\bibfnamefont {L.}~\bibnamefont {Pollet}},
  \bibinfo {author} {\bibfnamefont {F.}~\bibnamefont {Gerbier}}, \bibinfo
  {author} {\bibfnamefont {U.}~\bibnamefont {Schnorrberger}}, \bibinfo {author}
  {\bibfnamefont {I.}~\bibnamefont {Bloch}}, \bibinfo {author} {\bibfnamefont
  {N.~V.}\ \bibnamefont {Prokof'ev}}, \bibinfo {author} {\bibfnamefont
  {B.}~\bibnamefont {Svistunov}}, \ and\ \bibinfo {author} {\bibfnamefont
  {M.}~\bibnamefont {Troyer}},\ }\href {\doibase 10.1038/nphys1799} {\bibfield
  {journal} {\bibinfo  {journal} {Nature Physics}\ }\textbf {\bibinfo {volume}
  {6}},\ \bibinfo {pages} {998} (\bibinfo {year} {2010})}\BibitemShut {NoStop}%
\bibitem [{\citenamefont {Mark}\ \emph {et~al.}(2011)\citenamefont {Mark},
  \citenamefont {Haller}, \citenamefont {Lauber}, \citenamefont {Danzl},
  \citenamefont {Daley},\ and\ \citenamefont {N\"agerl}}]{mark2011}%
  \BibitemOpen
  \bibfield  {author} {\bibinfo {author} {\bibfnamefont {M.~J.}\ \bibnamefont
  {Mark}}, \bibinfo {author} {\bibfnamefont {E.}~\bibnamefont {Haller}},
  \bibinfo {author} {\bibfnamefont {K.}~\bibnamefont {Lauber}}, \bibinfo
  {author} {\bibfnamefont {J.~G.}\ \bibnamefont {Danzl}}, \bibinfo {author}
  {\bibfnamefont {A.~J.}\ \bibnamefont {Daley}}, \ and\ \bibinfo {author}
  {\bibfnamefont {H.-C.}\ \bibnamefont {N\"agerl}},\ }\href {\doibase
  10.1103/PhysRevLett.107.175301} {\bibfield  {journal} {\bibinfo  {journal}
  {Phys. Rev. Lett.}\ }\textbf {\bibinfo {volume} {107}},\ \bibinfo {pages}
  {175301} (\bibinfo {year} {2011})}\BibitemShut {NoStop}%
\bibitem [{\citenamefont {Thomas}\ \emph {et~al.}(2017)\citenamefont {Thomas},
  \citenamefont {Barter}, \citenamefont {Leung}, \citenamefont {Okano},
  \citenamefont {Jo}, \citenamefont {Guzman}, \citenamefont {Kimchi},
  \citenamefont {Vishwanath},\ and\ \citenamefont {Stamper-Kurn}}]{thomas2017}%
  \BibitemOpen
  \bibfield  {author} {\bibinfo {author} {\bibfnamefont {C.~K.}\ \bibnamefont
  {Thomas}}, \bibinfo {author} {\bibfnamefont {T.~H.}\ \bibnamefont {Barter}},
  \bibinfo {author} {\bibfnamefont {T.-H.}\ \bibnamefont {Leung}}, \bibinfo
  {author} {\bibfnamefont {M.}~\bibnamefont {Okano}}, \bibinfo {author}
  {\bibfnamefont {G.-B.}\ \bibnamefont {Jo}}, \bibinfo {author} {\bibfnamefont
  {J.}~\bibnamefont {Guzman}}, \bibinfo {author} {\bibfnamefont
  {I.}~\bibnamefont {Kimchi}}, \bibinfo {author} {\bibfnamefont
  {A.}~\bibnamefont {Vishwanath}}, \ and\ \bibinfo {author} {\bibfnamefont
  {D.~M.}\ \bibnamefont {Stamper-Kurn}},\ }\href {\doibase
  10.1103/PhysRevLett.119.100402} {\bibfield  {journal} {\bibinfo  {journal}
  {Phys. Rev. Lett.}\ }\textbf {\bibinfo {volume} {119}},\ \bibinfo {pages}
  {100402} (\bibinfo {year} {2017})}\BibitemShut {NoStop}%
\bibitem [{\citenamefont {Carcy}\ \emph {et~al.}(2021)\citenamefont {Carcy},
  \citenamefont {Herc\'e}, \citenamefont {Tenart}, \citenamefont {Roscilde},\
  and\ \citenamefont {Cl\'ement}}]{carcy2021}%
  \BibitemOpen
  \bibfield  {author} {\bibinfo {author} {\bibfnamefont {C.}~\bibnamefont
  {Carcy}}, \bibinfo {author} {\bibfnamefont {G.}~\bibnamefont {Herc\'e}},
  \bibinfo {author} {\bibfnamefont {A.}~\bibnamefont {Tenart}}, \bibinfo
  {author} {\bibfnamefont {T.}~\bibnamefont {Roscilde}}, \ and\ \bibinfo
  {author} {\bibfnamefont {D.}~\bibnamefont {Cl\'ement}},\ }\href {\doibase
  10.1103/PhysRevLett.126.045301} {\bibfield  {journal} {\bibinfo  {journal}
  {Phys. Rev. Lett.}\ }\textbf {\bibinfo {volume} {126}},\ \bibinfo {pages}
  {045301} (\bibinfo {year} {2021})}\BibitemShut {NoStop}%
\bibitem [{\citenamefont {Gerbier}\ \emph {et~al.}(2005)\citenamefont
  {Gerbier}, \citenamefont {Widera}, \citenamefont {F\"olling}, \citenamefont
  {Mandel}, \citenamefont {Gericke},\ and\ \citenamefont
  {Bloch}}]{gerbier2005}%
  \BibitemOpen
  \bibfield  {author} {\bibinfo {author} {\bibfnamefont {F.}~\bibnamefont
  {Gerbier}}, \bibinfo {author} {\bibfnamefont {A.}~\bibnamefont {Widera}},
  \bibinfo {author} {\bibfnamefont {S.}~\bibnamefont {F\"olling}}, \bibinfo
  {author} {\bibfnamefont {O.}~\bibnamefont {Mandel}}, \bibinfo {author}
  {\bibfnamefont {T.}~\bibnamefont {Gericke}}, \ and\ \bibinfo {author}
  {\bibfnamefont {I.}~\bibnamefont {Bloch}},\ }\href {\doibase
  10.1103/PhysRevLett.95.050404} {\bibfield  {journal} {\bibinfo  {journal}
  {Phys. Rev. Lett.}\ }\textbf {\bibinfo {volume} {95}},\ \bibinfo {pages}
  {050404} (\bibinfo {year} {2005})}\BibitemShut {NoStop}%
\bibitem [{\citenamefont {Capogrosso-Sansone}\ \emph
  {et~al.}(2007)\citenamefont {Capogrosso-Sansone}, \citenamefont {Prokof'ev},\
  and\ \citenamefont {Svistunov}}]{CapogrossoSansoneetal2007}%
  \BibitemOpen
  \bibfield  {author} {\bibinfo {author} {\bibfnamefont {B.}~\bibnamefont
  {Capogrosso-Sansone}}, \bibinfo {author} {\bibfnamefont {N.~V.}\ \bibnamefont
  {Prokof'ev}}, \ and\ \bibinfo {author} {\bibfnamefont {B.~V.}\ \bibnamefont
  {Svistunov}},\ }\href {\doibase 10.1103/PhysRevB.75.134302} {\bibfield
  {journal} {\bibinfo  {journal} {Phys. Rev. B}\ }\textbf {\bibinfo {volume}
  {75}},\ \bibinfo {pages} {134302} (\bibinfo {year} {2007})}\BibitemShut
  {NoStop}%
\bibitem [{\citenamefont {Cayla}\ \emph {et~al.}(2018)\citenamefont {Cayla},
  \citenamefont {Carcy}, \citenamefont {Bouton}, \citenamefont {Chang},
  \citenamefont {Carleo}, \citenamefont {Mancini},\ and\ \citenamefont
  {Cl\'ement}}]{cayla2018}%
  \BibitemOpen
  \bibfield  {author} {\bibinfo {author} {\bibfnamefont {H.}~\bibnamefont
  {Cayla}}, \bibinfo {author} {\bibfnamefont {C.}~\bibnamefont {Carcy}},
  \bibinfo {author} {\bibfnamefont {Q.}~\bibnamefont {Bouton}}, \bibinfo
  {author} {\bibfnamefont {R.}~\bibnamefont {Chang}}, \bibinfo {author}
  {\bibfnamefont {G.}~\bibnamefont {Carleo}}, \bibinfo {author} {\bibfnamefont
  {M.}~\bibnamefont {Mancini}}, \ and\ \bibinfo {author} {\bibfnamefont
  {D.}~\bibnamefont {Cl\'ement}},\ }\href {\doibase 10.1103/PhysRevA.97.061609}
  {\bibfield  {journal} {\bibinfo  {journal} {Phys. Rev. A}\ }\textbf {\bibinfo
  {volume} {97}},\ \bibinfo {pages} {061609} (\bibinfo {year}
  {2018})}\BibitemShut {NoStop}%
\bibitem [{\citenamefont {Tenart}\ \emph {et~al.}(2020)\citenamefont {Tenart},
  \citenamefont {Carcy}, \citenamefont {Cayla}, \citenamefont {Bourdel},
  \citenamefont {Mancini},\ and\ \citenamefont {Cl\'ement}}]{tenart2020}%
  \BibitemOpen
  \bibfield  {author} {\bibinfo {author} {\bibfnamefont {A.}~\bibnamefont
  {Tenart}}, \bibinfo {author} {\bibfnamefont {C.}~\bibnamefont {Carcy}},
  \bibinfo {author} {\bibfnamefont {H.}~\bibnamefont {Cayla}}, \bibinfo
  {author} {\bibfnamefont {T.}~\bibnamefont {Bourdel}}, \bibinfo {author}
  {\bibfnamefont {M.}~\bibnamefont {Mancini}}, \ and\ \bibinfo {author}
  {\bibfnamefont {D.}~\bibnamefont {Cl\'ement}},\ }\href {\doibase
  10.1103/PhysRevResearch.2.013017} {\bibfield  {journal} {\bibinfo  {journal}
  {Phys. Rev. Research}\ }\textbf {\bibinfo {volume} {2}},\ \bibinfo {pages}
  {013017} (\bibinfo {year} {2020})}\BibitemShut {NoStop}%
\bibitem [{\citenamefont {Ketterle.~W}\ and\ \citenamefont
  {Stamper-Kurn}(1998)}]{ketterle1998}%
  \BibitemOpen
  \bibfield  {author} {\bibinfo {author} {\bibfnamefont {D.~S.}\ \bibnamefont
  {Ketterle.~W}, \bibfnamefont {Durfee}}\ and\ \bibinfo {author} {\bibfnamefont
  {D.~M.}\ \bibnamefont {Stamper-Kurn}},\ }in\ \href@noop {} {\emph {\bibinfo
  {booktitle} {Bose-Einstein Condensation in Atomic Gases, Proceedings of the
  International School of Physics ''Enrico Fermi'', Course 140}}},\ \bibinfo
  {editor} {edited by\ \bibinfo {editor} {\bibfnamefont {S.~S.}\ \bibnamefont
  {Inguscio}, \bibfnamefont {M.}}\ and\ \bibinfo {editor} {\bibfnamefont
  {C.~E.}\ \bibnamefont {Wieman}}}\ (\bibinfo  {publisher} {IOS Press},\
  \bibinfo {address} {Amsterdam},\ \bibinfo {year} {1998})\BibitemShut
  {NoStop}%
\bibitem [{\citenamefont {Ma}\ \emph {et~al.}(2008)\citenamefont {Ma},
  \citenamefont {Yang}, \citenamefont {Pollet}, \citenamefont {Porto},
  \citenamefont {Troyer},\ and\ \citenamefont {Zhang}}]{ma2008}%
  \BibitemOpen
  \bibfield  {author} {\bibinfo {author} {\bibfnamefont {P.~N.}\ \bibnamefont
  {Ma}}, \bibinfo {author} {\bibfnamefont {K.~Y.}\ \bibnamefont {Yang}},
  \bibinfo {author} {\bibfnamefont {L.}~\bibnamefont {Pollet}}, \bibinfo
  {author} {\bibfnamefont {J.~V.}\ \bibnamefont {Porto}}, \bibinfo {author}
  {\bibfnamefont {M.}~\bibnamefont {Troyer}}, \ and\ \bibinfo {author}
  {\bibfnamefont {F.~C.}\ \bibnamefont {Zhang}},\ }\href {\doibase
  10.1103/PhysRevA.78.023605} {\bibfield  {journal} {\bibinfo  {journal} {Phys.
  Rev. A}\ }\textbf {\bibinfo {volume} {78}},\ \bibinfo {pages} {023605}
  (\bibinfo {year} {2008})}\BibitemShut {NoStop}%
\bibitem [{\citenamefont {Harada}\ and\ \citenamefont
  {Kawashima}(1998)}]{HaradaK1998}%
  \BibitemOpen
  \bibfield  {author} {\bibinfo {author} {\bibfnamefont {K.}~\bibnamefont
  {Harada}}\ and\ \bibinfo {author} {\bibfnamefont {N.}~\bibnamefont
  {Kawashima}},\ }\href {\doibase 10.1143/JPSJ.67.2768} {\bibfield  {journal}
  {\bibinfo  {journal} {Journal of the Physical Society of Japan}\ }\textbf
  {\bibinfo {volume} {67}},\ \bibinfo {pages} {2768} (\bibinfo {year}
  {1998})}\BibitemShut {NoStop}%
\bibitem [{\citenamefont {Campostrini}\ \emph {et~al.}(2001)\citenamefont
  {Campostrini}, \citenamefont {Hasenbusch}, \citenamefont {Pelissetto},
  \citenamefont {Rossi},\ and\ \citenamefont {Vicari}}]{campostrini2001}%
  \BibitemOpen
  \bibfield  {author} {\bibinfo {author} {\bibfnamefont {M.}~\bibnamefont
  {Campostrini}}, \bibinfo {author} {\bibfnamefont {M.}~\bibnamefont
  {Hasenbusch}}, \bibinfo {author} {\bibfnamefont {A.}~\bibnamefont
  {Pelissetto}}, \bibinfo {author} {\bibfnamefont {P.}~\bibnamefont {Rossi}}, \
  and\ \bibinfo {author} {\bibfnamefont {E.}~\bibnamefont {Vicari}},\ }\href
  {\doibase 10.1103/PhysRevB.63.214503} {\bibfield  {journal} {\bibinfo
  {journal} {Phys. Rev. B}\ }\textbf {\bibinfo {volume} {63}},\ \bibinfo
  {pages} {214503} (\bibinfo {year} {2001})}\BibitemShut {NoStop}%
\bibitem [{\citenamefont {Gaunt}\ \emph {et~al.}(2013)\citenamefont {Gaunt},
  \citenamefont {Schmidutz}, \citenamefont {Gotlibovych}, \citenamefont
  {Smith},\ and\ \citenamefont {Hadzibabic}}]{Gaunt2013}%
  \BibitemOpen
  \bibfield  {author} {\bibinfo {author} {\bibfnamefont {A.~L.}\ \bibnamefont
  {Gaunt}}, \bibinfo {author} {\bibfnamefont {T.~F.}\ \bibnamefont
  {Schmidutz}}, \bibinfo {author} {\bibfnamefont {I.}~\bibnamefont
  {Gotlibovych}}, \bibinfo {author} {\bibfnamefont {R.~P.}\ \bibnamefont
  {Smith}}, \ and\ \bibinfo {author} {\bibfnamefont {Z.}~\bibnamefont
  {Hadzibabic}},\ }\href {\doibase 10.1103/PhysRevLett.110.200406} {\bibfield
  {journal} {\bibinfo  {journal} {Phys. Rev. Lett.}\ }\textbf {\bibinfo
  {volume} {110}},\ \bibinfo {pages} {200406} (\bibinfo {year}
  {2013})}\BibitemShut {NoStop}%
\bibitem [{\citenamefont {Sagi}\ \emph {et~al.}(2012)\citenamefont {Sagi},
  \citenamefont {Drake}, \citenamefont {Paudel},\ and\ \citenamefont
  {Jin}}]{Sagietal2012}%
  \BibitemOpen
  \bibfield  {author} {\bibinfo {author} {\bibfnamefont {Y.}~\bibnamefont
  {Sagi}}, \bibinfo {author} {\bibfnamefont {T.~E.}\ \bibnamefont {Drake}},
  \bibinfo {author} {\bibfnamefont {R.}~\bibnamefont {Paudel}}, \ and\ \bibinfo
  {author} {\bibfnamefont {D.~S.}\ \bibnamefont {Jin}},\ }\href {\doibase
  10.1103/PhysRevLett.109.220402} {\bibfield  {journal} {\bibinfo  {journal}
  {Phys. Rev. Lett.}\ }\textbf {\bibinfo {volume} {109}},\ \bibinfo {pages}
  {220402} (\bibinfo {year} {2012})}\BibitemShut {NoStop}%
\end{thebibliography}%


\end{document}